# Computer Supported Collaborative Processes in Virtual Organizations

Zbigniew Paszkiewicz, Poznań University of Economics, Poland
Wojciech Cellary, Poznań University of Economics, Poland

*In global economy, turbulent organization environment strongly influences organization's operation. Organizations must constantly adapt to changing circumstances and search for new possibilities of gaining competitive advantage. To face this challenge, small organizations base their operation on collaboration within Virtual Organizations (VOs). VO operation is based on collaborative processes. Due to dynamism and required flexibility of collaborative processes, existing business information systems are insufficient to efficiently support them. In this paper a novel method for supporting collaborative processes based on process mining techniques is proposed. The method allows activity patterns in various instances of collaborative processes to be identified and used for recommendation of activities. This provides an opportunity for better computer support of collaborative processes leading to more efficient and effective realization of business goals.*

## Introduction

Organization environment is defined as "all the forces, processes and other entities – companies, public administration agencies, non-government organizations, etc. – outside an organization that interact with the organization and can potentially affect the organization's performance" (Stoner, Freeman, & Gilbert, 1999). In global economy, organization environment strongly influences organization's operation and its market success. Current trends: globalization, development and proliferation of information technology, development of knowledge-based economy and rising competition, result in increased complexity, uncertainty, dynamism, turbulence and diversity of organization environment. Such environment is a particular challenge for small organizations. Although small organizations are flexible and innovative and they adapt to changing environment in a relatively easy and fast way, they are volatile, have limited capabilities to influence the market and to control their environment and, finally, compete with large global organizations that have much more resources. To face this challenge, small organizations base their operation on strategies of specialization, differentiation and collaboration within Virtual Organizations.

The concept of a *Virtual Organization (VO)* (Camarinha-Matos, Afsarmanesh, & Ollus, 2008) was proposed as an approach to cooperation among multiple autonomous partners – organizations, humans or information systems – cooperating via Internet and strongly supported by information technologies. The main challenge of virtual organizations is an efficient collaboration of autonomous partners to achieve a predefined goal, and, if needed, to quickly adapt to changing environment. Such adaptation helps to reduce business risk or to take advantage of new business opportunities. Theoretical foundations for virtual organizations have been proposed in (Camarinha-Matos, Afsarmanesh, & Ollus, 2008) and (Camarinha-Matos, Afsarmanesh, Galeano, & Molina, 2009).

VO operation requires new IT techniques to support VO processes. Existing approaches in this matter are based on two basic assumptions: (1) the process model is known before process execution, i.e., it is possible to design a process model that is further instantiated and later on executed; (2) the business environment is rather static, which implies process repetition. Processes meeting the above two assumptions are fully structured, repeatable and predictable. Examples of such processes are production ones which can be well supported. To support such processes, a number of methods and standards have been proposed by Russell and Aalst (2007) and in WS-BPEL (2011), WS-Coordination (2011), WS-Choreography (2011) standards. However, full automation is achieved at the expense of adaptation which cannot be made during process runtime. In case of organizations operating in dynamic environments, the two above assumptions are not necessarily observed. If a business environment is highly dynamic, it may not be possible to foresee the process that has to be performed in a given moment. Such processes have often ad-hoc character, so the IT support should focus on communication among process participants (Swenson, 2010) and flexible re-definition of process model during runtime (Sadiq, Orłowska, Sadiq, 2005). The goal of IT support is to maximize adaptability of activities performed by organizations to changing circumstances. In practice, collaborative processes of virtual organizations are often semi-structured. On the one hand, collaborative processes are highly unpredictable, but on the other hand, it is possible to distinguish a set of activities that frequently appear in a particular context (Swenson, 2010). This means they require the mix of both above approaches to be applied. Methods of support for such processes are still to be developed.

The main contribution of this paper is: (1) a concept of application of process mining techniques to support collaborative processes during their execution, (2) a novel method for identification and recommendation of *activity patterns* based on process mining techniques, where an activity pattern is a set of activities that are frequently performed in a structured way in a particular context. The proposed method has been partially implemented in the *ErGo* system (2011) developed within the ITSOA project (2011).

The remainder of this paper is organized as follows. In Section 2, the concept of virtual organization is presented in detail. Moreover, the concept of Virtual Organization Breeding Environment as an organizational approach to facilitate operation of virtual organizations is described. In Section 3, basic notions concerning process modeling, analysis and execution of processes are presented. In particular, the concept of semi-structured and collaborative processes is described. In Section 4, a method supporting execution of semi-structured collaborative processes based on process mining concept with an illustrative example is presented. Finally, Section 5 concludes the paper.

## Virtual organizations

To achieve market success, small organizations have to combine strategies of specialization, differentiation and cost leadership throughout the value chain. In consequence, these strategies lead to integration of various organization efforts and cooperation.

The main reason for collaboration among organizations is the need for *competitive advantage*. The first theoretical framework that may be used to understand the fundamental need for collaboration among organizations has been proposed by David Ricardo (1817) in his book "Principles of Political Economy and Taxation". David Ricardo indicated the strategy of *specialization* as a way to boost efficiency of organization operation. Specialization means concentration of an organization on operations where it has comparative advantage and taking benefits from exchange of goods and services with other specialized organizations. Ricardo has explained that this approach is effective even if an organization is able to produce all the goods and services more efficiently than the other organizations.

In the theory of *competitive advantage*, Michael Porter (1985) proposed a value chain as an approach to analysis of organization operations. A *value chain* is a set of activities related to production processes, marketing, supply, client support, etc. which all together lead to service provision or product delivery that has a value to a final customer. Basing on the value chain concept, two organization strategies were proposed: cost leadership and differentiation. An organization develops a *cost advantage* by reconfiguring its value chain to reduce costs of as many stages of the chain as possible. Reconfiguration means making structural changes, such as adding new production processes, changing distribution channels, or trying a different sales approach. *Differentiation* stems from uniqueness and perceived value. An organization focusing on activities it does best and creating innovative and unique products and services, naturally rises above its competitors. An organization can achieve a differentiation advantage by either changing individual value chain activities to increase uniqueness in the final product or by reconfiguring the entire value chain.

Focusing on the organization operation areas which provide the competitive advantage should be a fundamental aspect of every business strategy. This approach leads to narrowing the areas of organization expertise and operation. Meanwhile, the production and service provision currently require a large set of skills and resources that a given organization is usually not able to handle efficiently. Thus, modern value chains cover not one but a number of specialized organizations, integrated with each other to perform activities defined within the value chain. Such cooperation creates an opportunity for efficient, cost effective differentiation at each phase of the value chain.

The cooperation of specialized organizations within a single value chain allows small organizations to compete on the global market with large multinational organizations. Such cooperation is difficult mainly due to differences existing among autonomous organizations such as geographic, legislative, cultural differences, diverse markets of products and services, constant changeability of organization customers, suppliers, changeability of technology and methods of work, differences in used information systems, work organization and law, etc. Therefore, modern organizations seek for methods, techniques, strategies and organizational structures that facilitate development of efficient inter-organization cooperation. In this context the concept of a *Virtual Organization (VO)* was proposed by Camarinha-Matos, Afsarmanesh and Ollus (2008) and Camarinha-Matos, Afsarmanesh, Galeano and Molina (2009).

*Virtual Organization (VO)* is a set of at least two autonomous partners, where at least one of them is an organization, cooperating within a particular structure of social and legal relations in order to carry out a particular venture due to the demand from virtual organization clients and having a plan to carry out this venture. *Partners* collaborating within a VO are organizations – enterprises, public administration units, non-government organizations – people and information systems. The inherent, desired characteristics of a virtual organization are its adaptation to changing circumstances and constant learning.

A VO is a particular example of an adaptive system and follows its characteristics. An *adaptive system* is "set of elements which interact with each other and has at least one process which controls the system's adaptation, that is, the correlation between its structure, function or behavior and its environment, to increase its efficiency to achieve its goals" (José, Lope, & Maravall, 2009). In case of a VO, adaptation may be triggered either by external events occurring in VO political, social, economic or technological environment or internal events that are strictly connected to the execution of VO operational plan. In particular, adaptation may result in a modification of a set of VO partners, modification of operational plan or even a redefinition of the VO goal. Taking into account the characteristics of organization environment operating in global economy, it is common that an adaptation of virtual organization takes place during the whole VO operation.

Virtual organization being a learning organization (Argyris, 1999) constantly gains and maintains knowledge on aspects that may directly or indirectly influence its operation. Knowledge management in VO encompasses the knowledge of VO as a whole as well as the knowledge of particular VO partners. This knowledge includes information about appearing new products and processes in the environment, information concerning competences and services provided by current and potential partners, information about social relations among partners and other organizations, information on effectiveness of partners operation, specialized and domain specific knowledge on technologies and methods for provision of services and production of products etc. This knowledge is used for a proper organization of VO operation and production of creative solutions by all the partners being VO members.

Efficient cooperation requires the use of appropriate management strategies, techniques and structures, including: outsourcing and out-tasking strategies, techniques for efficient control of activity execution, standardization of non-critical areas of operation and interorganizational integration on various levels of organizational structure. Cooperation also forces new approaches to traditional areas of management such as evaluation of organization performance that is different when evaluating a set of collaborating organizations or a single, standalone organization. Information technology creates a foundation for VO operation and execution of VO strategies and techniques. Information technology enables efficient planning of operation, communication and coordination of actions, integration of partners, control of activity execution, measurement of business effectiveness etc.

The concept of *Virtual Organization Breeding Environment* (VOBE, sometimes referred to in the literature as VBE) has been proposed to facilitate VO operation. A VOBE is "an association of organizations with the main goal of increasing preparedness of its members towards collaboration in potential virtual organizations" (Camarinha-Matos, Afsarmanesh, & Ollus, 2008). VOBE allows potential collaborators to prepare their future collaboration with other VOBE members before a business opportunity occurs. It is possible to distinguish various types of VOBE including technology clusters or industry areas (Camarinha-Matos, Afsarmanesh, Galeano, & Molina, 2009). VOBE supports its members by providing an access to various sources of data and tools that can be used across virtual organization lifecycle (Picard, Paszkiewicz, Gabryszak, Krysztofiak, & Cellary, 2010):

- in the creation phase: VOBE provides access to information not publicly available, such as information about the past performance of VOBE members; it also provides a standardized description of partner profiles, competences and services; it supports the potential partner search and selection; it provides methods and tools for analysis and evaluation of present and future cooperation performance, as well as necessary information for trust building among selected members;
- in the operation phase: VOBE supports communication and exchange of documents, facilitates integration of heterogeneous information systems and manages common infrastructure, provides guidelines for standardized data formats, data storage facilities, information about changing environment (context) of collaboration, information about new collaboration opportunities, and permits to reuse artifacts elaborated by other VOs (in particular business process models, best practices);
- in the evolution phase: VOBE supports adaptation by redefinition of business goals, searching for new partners, supporting negotiations, etc;
- in the dissolution phase: VOBE inherits knowledge, i.e., it captures experience gained during the operation of VOs for future reuse.

# Collaborative processes in virtual organizations

The operation of a virtual organization is based on execution of collaborative processes. A collaborative process is a representation of a VO operational plan. Due to the specific characteristic of virtual organization and the characteristic of the environment a collaborative processes takes place, the VO collaborative processes create new challenges for its efficient execution and support with the use of information technologies.

## Basic concepts

An *activity* is a closed piece of work (WfMC, 1999). It may be a piece of automated work performed by an information system, e.g., a web service for creating invoices, a piece of work performed by a human, e.g., making a decision by a senior executive, or a piece of work performed by an organization, e.g. building a shopping mall. A set of activities which realize a business objective in a structured manner is called a *process* (WfMC, 1999). Information systems, humans and organizations being involved in activities being a part of the process are called *actors*.

Modern approaches to the definition of business processes are built around the concept of service. A *service* is an access to a competence of an actor, called *service provider*, to satisfy a need of a second actor, called *service consumer*, where the access is provided via a prescribed *interface*. In the approach used in this paper, each activity being a part of a process structure is performed by a *process participant* being a service consumer of a service provided by service provider. A *process element* may refer to a process participant, a service or a service provider. Finally, a *process instance* is a single enactment of a process that normally takes place within the context of an organizational structure defining functional roles of involved actors and relationships existing among actors. A process instance is executed in a particular *process context* that is a set of elements describing the process instance, executed activities and the environment it is executed in. A *state* is a representation of the internal conditions defining the status of a process instance at a particular moment. A *process model* captures the possibility to execute a given activity in a given state.

## Approaches to process execution support

In current approaches to support of process execution, one can indicate the mismatch between a classical business process management approach and highly dynamic business environments. The indication of this mismatch leads to distinguishing three types of processes: (1) structured, (2) ad-hoc and (3) semi-structured.

In a classical business process management approach, usually structured processes are considered. Two basic assumptions are made in this approach: (1) the process model is known a priori, i.e., it is possible to design a process model that is further instantiated and later on executed; (2) the business environment is rather static, which implies potential process repetition, i.e., a given process model may rule many process instances. This leads to a situation when the full definition of the process is known in advance including the network of activities and their ordering relationships, criteria to indicate the start and termination of the process, and information about the individual activities, such as participants, associated IT applications and data. Support for execution of fully structured processes is relatively easy due to its predictability and repeatability. IT support for the execution of such a structured processes takes a form of BPEL (2011), WS-Coordination (2011), WS-Choreography (2011) standards concerning automatic execution of business processes and focusing on the IT infrastructure layer.

In case of organizations operating in a highly dynamic environment these two assumptions are not necessarily observed. If a business environment is highly dynamic, where organizations are continuously adapting by modification of plans or goals, it may not be possible to foresee the process that will be performed. In such environment, it is difficult to predict both the set of activities that will be a part of the process and a set of required resources. Due to the fact that a process cannot be modeled a priori, the design of a process model should be performed in an ad-hoc manner at run-time. Missing process model makes supporting processes execution a difficult task. Therefore, IT support has to focus on communication among process participants (Swenson, 2010) and on approaches to flexible re-definition of process model during runtime (Sadiq, Orłowska, Sadiq, 2005).

The term *semi-structured process* refers to a process that is only partially structured in advance. The foundation for the concept of semi-structured processes was created by Swenson (2010) on a basis of knowledge work analysis, but he did not propose the term itself. Swenson has identified the problematic case of processes related with knowledge work, e.g., emergency rescue, financial audit or bridge construction engineering, for which he argues that a new approach is needed, referred to as *adaptive case management* (ACM). Among others, the main characteristic of such processes is their emergent aspect. The emergence refers to the influence of the execution of a process on the process itself. The process is unpredictable and evolves as subsequent actions take place. In the context of emergent processes, another observation can be made. Namely, although it is impossible to predict the full course of process execution, it can be noted that some sequences of actions are highly probable to appear in a particular context. For instance, consider a process describing a rescue action performed by a firefighter. The firefighter does not know how a particular rescue action will develop, but he/she is trained to behave in a certain way in a particular context. For example, when an electrical installation is on fire, he/she performs a known set of activities related with this situation. Semi-structured processes are modeled in an ad-hoc manner but in some context the structure is known *a priori*. Still, it is unknown whether the particular context will appear. The IT support for this kind of processes is not yet provided.

In the context of virtual organizations, among all semi-structured processes, one may distinguish a group of collaborative processes. A *collaborative process* is a semi-structured process that involves at least two different and autonomous process participants being legally independent and aiming to fulfill their own goals that may be different from the goal of a virtual organization they are a part of. The characteristics of the collaborative process are directly connected to adaptive and learning nature of a VO.

Adaptation of a VO to its complex, uncertain and turbulent environment results in the three following main characteristics of the VO collaborative process: (1) dynamism – the process evolves, where an evolution "may be slight as for process improvements or drastic as for process innovation or process reengineering. In any case, the assumption is that the process has pre-defined models, and business process change causes these models to be changed" (Sadiq, Orłowska, Sadiq, 2005) (2) flexibility – execution of the process starts without its full specification, i.e., the full set of activities to be performed and their ordering is not known when the process execution starts, so specification of the model is made at runtime and may be unique to each process instance; flexible processes are characterized by a lack of ability to completely predict and define a set of activities and ordering relationships among them; (3) adaptability – execution of the process adjusts to exceptional circumstances that may or may not be foreseen, and generally would affect one or a few process instances.

VO collaborative processes are knowledge intensive meaning that: (1) process involves a set of usually interdisciplinary, cross organizational teams comprised of members being highly qualified, specialized and having experience needed for performing a set of activities aiming to solve a complex problem; (2) process participants constantly gain a new knowledge through the analysis of information concerning the process context; moreover, partners learn from each other both the explicit and tacit rules governing the execution of the process; (3) as a consequence of the instantly gained knowledge, process participants change the way they perceive the process, activities and the semantics of the decisions being made and the way these decisions are made; (4) due to the long-lasting character of the collaborative processes, a set of collaborating partners and their roles change which results in the fact that a set of partners having the holistic vision and understanding of the process may be small; (5) similar instances of collaborative processes – e.g. having a similar goal, involving a similar group of participants, performed at the same time – may be interrelated, meaning that the course of execution of one process and its result may influence the course of execution of another instance. All these characteristics result in the fact that it is possible to predict and model only the generalized structure of the collaborative process but a detailed prediction of the process structure and execution is impossible.

# IT support for execution of collaborative processes

Efficient IT support for collaborative processes is still to be provided. Such processes require a mix of approaches to both structured and ad-hoc processes support. The key problem is to identify parts of a collaborative process that have a predictable structure and those which have ad-hoc character. Automation of the structured parts leads to improvement of execution of the whole collaborative processes.

An approach proposed in this paper to supporting collaborative processes is based on recommendation for activity patterns. An *activity pattern* is a set of activities that are frequently performed in a structured way in a particular context. The main idea is to analyze running and former executions of the collaborative process instances to detect activity patterns. Knowledge of activity patterns and the context they appear in can be used in recommendations. This approach requires the elaboration of techniques and methods for (1) identification of activity patterns in processes, (2) identification and analysis of activity pattern context, (3) recommendation for detected activity patterns in on-going and future processes.

It is possible to identify the activity patterns and their context explicitly. In the example presented in Section 3.2, a firefighter is explicitly taught during his/her training how to behave in particular situations. On the other hand, very often identification and usage of activity patterns is a matter of experience and tacit knowledge that is difficult to verbalize. Moreover, in case of processes performed by organizations it is easy to notice the discrepancies between the way the processes were intended and are executed in practice. Thus, a method for objective and human error free identification of activity patterns should be used. In this paper the concept of process mining is proposed to be used for this purpose.

## *Process mining*

The term process mining is used for "the method of distilling a structured process description from a set of real process executions" (Weijters, & Aalst, 2001). The concept of process mining is based on observation that creating a process modeling, as stated in a classical business process management approach, is a complicated and time-consuming process and typically there are discrepancies between the actually executed processes and the envisioned process models. Moreover, there are processes which cannot be modeled due to their unpredictability, but once they are executed, the knowledge concerning the model of executed process is still useful. In both cases, it is possible to use process mining techniques. Traditionally, in the research area of process mining, one may distinguish among others two main areas of interest, i.e. *process discovery* and *process conformance checking*.

Discovery of process models is based on exploration of *events* generated by executed process instances. Generated events are recorded and stored in an *event log*. The following assumptions concerning events are made (Weijters, & Aalst, 2001): (i) each event refers to a process activity, (ii) each event refers to a case (i.e., a process instance), and (iii) events are totally ordered (i.e., in the log, events are recorded sequentially even though tasks may be executed in parallel). Note that any information system used to support processes such as ERP, CRM, or workflow management systems offers this information in some form. These process logs are used to construct a process model of the behavior registered. Process conformance checking allows discovered process models to be compared with predefined models. The characteristics of processes that can be discovered is presented in (Aalst, Weijters, & Maruster, 2004) and (Aalst, & Weijters, 2004). Process conformance checking leads to process model improvement or extension and can trigger Business Process Reengineering (BPR) efforts or reconfiguration of information systems supporting information processes.

Recently, stronger emphasis has been put on using process mining at the process runtime. In this context the methods for process *prediction*, *checking* and *recommendation* have been proposed. Currently existing recommendation methods aim at prediction and advise the activities that will appear in a particular process instance in the nearest future, where recommendation is guided by the defined process participant's goal. For instance, the preference to finish the process as soon as possible or as cheap as possible (Aalst, Pesic, & Song, 2010). Current recommendation methods based on process mining techniques miss significant aspects including: internal and external process and activity context, understanding of semantics of particular activities and data connected to these activities, analysis of ongoing process instances, aspect of collaboration and flexible definition of process participant preferences concerning recommendations. Moreover, goals guiding recommendations cannot be defined on a business level, because the analysis of correlation among process instance structure and successful or unsuccessful process end is not performed. So far process recommendations were not analyzed in the context of activity patterns and collaborative processes.

### RMV method

The main idea of Recommendation Method for Virtual Organizations (RMV) is an automatic *ad-hoc* generation of recommendations for collaborative processes, where the collaborative process is performed within a virtual organization breeding environment.

The main method assumptions are: (1) a process that is supported is a collaborative process meaning that the detailed model of the process is not known in advance; (2) process participants use tools and data sources offered by VOBE for activity execution, however the set of tools and data sources changes if necessary as the process evolves; (3) the VOBE's technical infrastructure offers a possibility to record events created by process participant actions during activity execution, as well as the context of these activities; (4) process mining techniques are used for discovery of process instance models that are later analyzed in terms of activity patterns identification; (5) main phases of a supported collaborative process are known in advance.

The RMV method consists of two main phases (**Figure 1**): (1) activity pattern extraction phase, (2) recommendation formulation phase.

In the first phase, actions performed by users (step 1 in Figure 1) create system events that are logged in the event log (2). Event log stores information concerning completed and partially executed ongoing process instances. On the basis of this information, the models of particular process instances are discovered (3) and analyzed in terms of activity patterns (4). Identified activity patterns together with their context are stored in the repository (5).

In the second phase, recommendation is performed as a response to a request (6). The request must include information concerning the trace of process execution to date and current process context. On the basis of information available in the request, activity patterns are searched (7) and best fitting activity patterns are used for the recommendation (8). The recommendation is a set of triples: (a) activity pattern; (b) recommendation confidence indicator stating how well a particular activity pattern matches the current process context; (c) optional justification of recommendation.

The important aspects of the method are: parameterization, four levels of analysis of completed and ongoing processes and context analysis.

The recommendation mechanism in the RMV method can be parameterized to meet user-specific expectations. Parameterization is done in the form of user preferences. It concerns various aspects of recommendation mechanism, for instance: definition of the context that should be taken into account during process analysis, balance between user-centric and crowd-based recommendations, minimum number of activities being a part of an activity pattern, etc.

The analysis of processes and activity templates performed during activity pattern identification and search is done on four levels: (1) process structure, (2) process activities and data associated with these activities, (3) process participants associated with activities, (4) process instance and activity context.

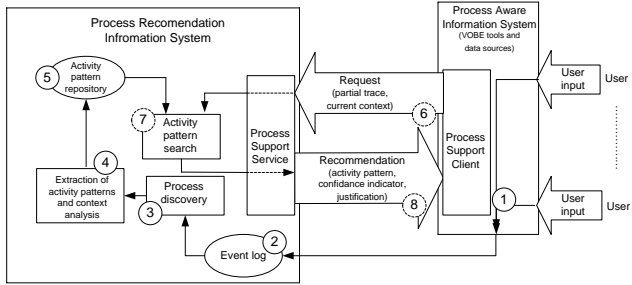

Figure 1. RMV method steps and Process Recommendation Information System infrastructure. Numbers in circles with solid borders (1-5) correspond to phase 1 of the method. Dashed circles (steps 6-8) comprise the phase 2.

These four levels allow activity patterns and processes to be appropriately categorized and compared. Both completed and ongoing processes are analyzed to detect activity patterns. However, during recommendation phase only activity patterns identified in successful processes are taken into consideration. Note that a definition of what is a successful process is a part of user preferences, for instance some users may prefer short process executions, meanwhile other may prefer the opposite. The set of activity patterns analyzed for recommendation is different in each case.

Two types of process context are distinguished: internal and external ones. Internal context concerns the data directly associated with the process and its activities, for instance the time of activity execution, or tools and data used for performing a particular activity. External context describes the state of social, economic and political environment a collaborative process takes place in. The scope of context information useful for a particular collaborative process depends strongly on its nature and can be modified following user preferences. Inclusion of internal context in reasoning and recommendation mechanism makes it possible to generate recommendations concerning activity parameters, e.g, it is possible to recommend a person to perform the next activity basing on information about who performed the previous one.

As mentioned above, all the recommended processes are performed within the VOBE. The tools and data sources available in VOBE, no matter whether they are used in a particular collaborative process by process participants, provide necessary data for context definition. Useful information for this purpose may concern for instance the current status of social relations existing among organizations in VOBE.

The RMV method encompasses: (1) extraction of activity patterns from event log and their analysis; (2) comparison of process instances and activity templates structures and context; (3) creation of recommendations based on activity pattern repository, information about the external and internal context and current process instance;

(4) parameterization of recommendation mechanism based on user preferences.

The advantage of identification of activity patterns and recommendation basing on process mining techniques is the ability to catch the tacit knowledge of collaborative process participants. Due to the analysis of data concerning all activities performed by collaborative process participants which are available in event log, it is possible to identify how people actually perform processes within a virtual organization. This minimizes the risk of process and process model discrepancies. The recommendation information can be used by a Process Support Client (**Figure 1**) in various ways. Simple scenario may concern information being simply displayed to a user for his/her recognition. More advanced applications can concern passing the information to workflow engine that supports flexible definition of workflow processes. Recommendations of activity patterns enable boosting efficiency of collaborative processes by providing computer support specific to structured processes on those parts of the process where activity patterns have been identified and matched to process context.

### Technical implementation

The proposed method has been partially implemented in the *ErGo* system developed within the ITSOA project. The *ErGo* system is designed as an IT infrastructure supporting operation of virtual organization breeding environment. The platform encompasses a set of tools and data sources that can be offered for the use by VOBE members. The *ErGo* system is envisioned to be used in the Polish constructing sector.

### Illustrative example

The VO collaborative process aims at delivery of products or provision of services to VO clients. This process is accompanied by other collaborative processes having the supporting role. An example of such supporting collaborative process is partner selection (Paszkiewicz, & Picard, 2010) performed throughout the VO operation. Let us consider the use of RMV method for finding the appropriate single partner for the VO collaborative process.

The RMV method is first parameterized. In this example, parameterization includes the two following facts: (1) mining rule stating that only those activity sequences that appear in at least three process instances can be considered as activity patterns, (2) activity internal context includes data concerning: process participants, tools and data used by the tools. The analyzed process is performed by five participants: P1, P2, P3, P4, P5 (step 1 of the RMV method, *cf.* Figure 1). Actions are recorded and saved in the event log (step 2). Process instances executed by the process participants are presented in Figure 2 – six executions of the process are represented by two process instance models. Each activity (white rectangle, *cf.* Figure 2) is recorded with a corresponding internal context: process participants that performed the activity (grey rectangle), VOBE tool used (rectangle with text in italics) and data used in activity (rectangle with dashed border). For instance, *partner search* in process instance model *a* in Figure 2 was performed by P2 and search engine and localization criteria were used during the search. For clarity, in Figure 3 the activity contexts are not presented. It is assumed that all six process instances occurred in the same *external context c1*.

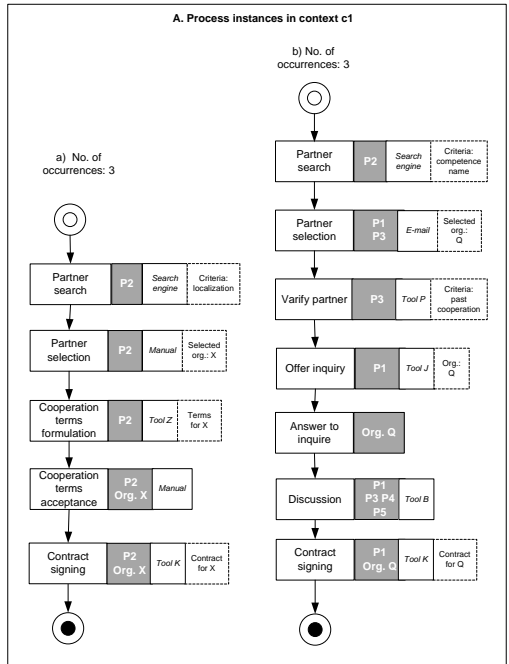

Figure 2. Partner selection process instances performed in context c1.

The process model (Figure 3B) is extracted from the recorded data (step 3). The extracted process model encompasses models of all the recorded instances and is further analyzed to detect (step 4) and store (step 5) activity patterns (Figure 4). Detected activity patterns, presented in Figure 4, indicate that: (a) it is common to start the process with two activities: *partner search* and *partner selection*; (b) if the *partner selection* is performed *manually*, it is followed by *formulation of cooperation terms*, *cooperation terms agreement* and *contract signing;* (c) if *partner selection* is done with the use of *email*, the activities from activity pattern *c)* take place. Activity patterns encompass information about process participants, tools and data, for instance *formulation of cooperation terms* is usually executed by *P2*, with the use of *tool Z* with data concerning *organization X*. Note that in Figure 4 only maximal activity patterns are presented – each subsequence of activities creating a maximal activity pattern is also an activity pattern.

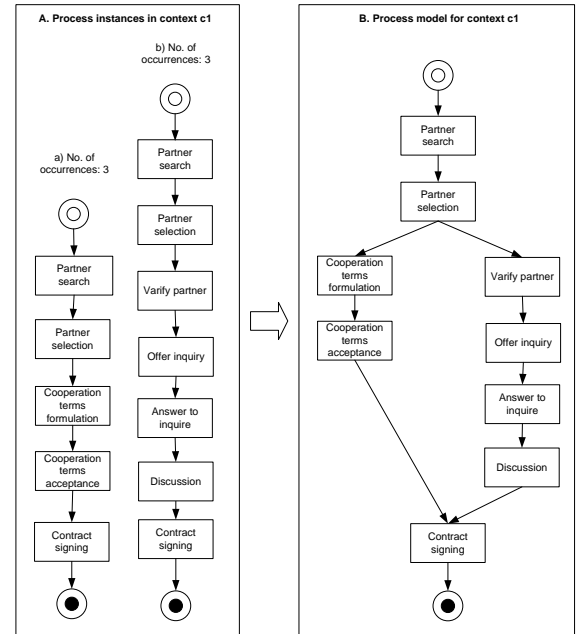

Figure 3. RMV method outcomes for analysis of process performed in context c1: A. process instances, B. Process model.

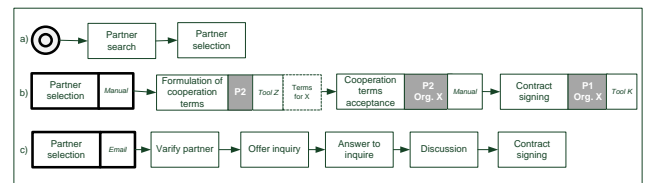

Figure 4. Activity patterns for context c1.

The external context of the process changes in time. In Figure 5A a process model extracted on a basis of process instances that took place in context c2 is presented. Activity pattern extracted from the process model is shown in Figure 5B. Note that *partner verification* activity is not a part of the activity pattern as it does not meet the requirement defined by a process participant during method parameterization (minimum three occurrences).

The recommendation mechanism is triggered by a process participant request. Consider a process participant who sends an email informing that he/she selected a partner in the context *c1*. Three identified earlier activity patterns are analyzed in terms of meeting participant process characteristics (step 7). Note that the occurrence of partner selection does not determine clearly which activity pattern to use. The determination of process execution path is based on the internal context– *manual* execution of *partner selection* determines the use of activity pattern *b*, the use of *email* – activity pattern *c*.

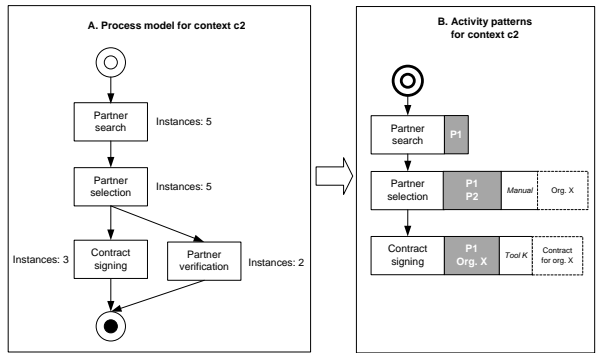

Figure 5. RMV method outcomes for analysis of process performed in context c2: A. Process model, B. Activity patterns.

The user receives the following recommendation (step 8):

- <<verify partner (P3, tool P, Criteria: past cooperation), offer inquiry (P1, Tool Z, Organization Q), answer to inquiry (Organization Q), discussion ((P1, P3, P4, P5), Tool K), contract signing (P1, Organization Q, Tool K, Contract for Q)>, confidence indicator: 100%>
- <<formulation of cooperation terms (P2, Tool Z, Terms for X), cooperation terms acceptance (P2, Organization X), contract signing (P1, Organization X)>, confidence indicator: 75%>

If his/her process were performed in context similar to *c2*, the following recommendation would be formulated:

- <<contract signing ((P1, Org. X), Tool K, Contract for org. X)>, confidence indicator: 75%>

Confidence indicator is not maximal if the external context or the internal activity context does not fully match the context of an activity pattern.

## Conclusions

As follows from this paper, the RMV method based on process mining techniques to support collaborative processes during their execution may produce valuable recommendations. It has been proved that historical data generated by completed collaborative process instances may be efficiently used to support ongoing instances in a form of recommendations. The created recommendations are adjusted to the external and internal context of the ongoing process which makes them more accurate and applicable. Moreover, context analysis supports the adaptation of the collaborative processes to changing environment. When the context changes, the set of recommended actions changes as well. Inclusion of the internal context makes it possible to generate recommendations concerning activity parameters. The data concerning process context are provided by VOBE.

Identification and recommendation of activity patterns is justified in the context of collaborative processes. For such processes, it is possible to recommend not only one activity, as in currently available methods, but a set of activities that usually appear together. The context in which an activity pattern appeared in the past is taken into account during its selection for recommendation. Applicability of a recommendation in the current process context is measured and indicated by the confidence indicator.

By taking advantage of context analysis and activity patterns, the RMV method contributes to adaptability and efficiency of collaborative process execution by identification of its repeatable parts and by creating opportunity for their support.

Further development of the RMV method is planned within the IT-SOA project. In particular, efficiency of algorithms responsible for identification and analysis of activity patterns and their context should be verified. The RMV method verification is planned with a pilot application in the construction sector.

**Acknowledgments.** This work has been partially supported by the Polish Ministry of Science and Higher Education within the European Regional Development Fund, Grant No. POIG.01.03.01-00-008/08.